\documentclass[aps,prl,nofootinbib,amsmath,amssymb,superscriptaddress,twocolumn,10pt]{revtex4}%superscriptaddress,showpacs,

\pdfoutput=1

\usepackage{graphicx}
\usepackage{dcolumn}
\usepackage{bm}
\usepackage{amssymb}
\usepackage{latexsym}
\usepackage{booktabs}
\usepackage{amsmath}
\usepackage{multirow}
\usepackage[colorlinks=true, linkcolor=red, citecolor=blue]{hyperref}

\newcommand{\be}{\begin{equation}}
\newcommand{\ee}{\end{equation}}
\newcommand{\bq}{\begin{eqnarray}}
\newcommand{\eq}{\end{eqnarray}}

\usepackage[usenames,dvipsnames]{xcolor}

\begin{document}

\title{Cosmological parameter measurement and neutral hydrogen 21 cm sky survey with the Square Kilometre Array}

\author{Yidong Xu}
\email{xuyd@nao.cas.cn}
\affiliation{Key Laboratory for Computational Astrophysics, National Astronomical Observatories, Chinese Academy of Sciences, Beijing 100101, China}
\author{Xin Zhang}%\footnote{Corresponding author}}
\email{zhangxin@mail.neu.edu.cn}
\affiliation{Department of Physics, College of Sciences, Northeastern
University, Shenyang 110819, China} 
%\affiliation{Ministry of Education's Key Laboratory of Data Analytics and Optimization for Smart Industry, Northeastern University, Shenyang 110819, China}
%\affiliation{Center for High Energy Physics, Peking University, Beijing 100080, China}
%\affiliation{Center for Gravitation and Cosmology, Yangzhou University, Yangzhou 225009, China}

\begin{abstract}

In order to precisely measure the cosmological parameters and answer the fundamental questions in cosmology, it is necessary to develop new, powerful cosmological probes, in addition to the proposed next-generation optical survey projects. The neutral hydrogen 21 cm sky surveys will  provide a promising tool to study the late-universe evolution, helping shed light on the nature of dark energy. The Square Kilometre Array is the largest radio telescope in the world to be constructed in the near future, and it will push the 21 cm cosmology into a new era and greatly promote the development of cosmology in the forthcoming decades.

\end{abstract}
%\pacs{95.36.+x, 98.80.Es, 98.80.-k}
\maketitle

%Cosmologists have long dreamed of a standard model of cosmology. 
Through decades of continuous efforts by the whole community of cosmology, 
a standard cosmological model known as the $\Lambda$ cold dark matter ($\Lambda$CDM) model has been established, 
in which the total energy budget consists of 5\% ordinary matter, 27\% cold dark matter, and 68\% dark energy 
described by a cosmological constant $\Lambda$, and the inflation process in the very early universe yielded 
adiabatic, Gaussian, nearly scale-invariant primordial density perturbations, and possibly detectable primordial gravitational waves.
The $\Lambda$CDM model with several ansatzes can fit the observational data of cosmology with breathtaking precision. Actually, 
the Planck satellite mission has implemented unprecedentedly precise measurements for the cosmic microwave background (CMB) anisotropies, 
which has pushed the studies of the cosmos into the era of ``precision cosmology''. 
The latest observational results of the CMB power spectra strongly favor a ``base'' version of the $\Lambda$CDM model that has only six basic cosmological parameters and a number of well-tested ansatzes. 

%The constraints on the base-$\Lambda$CDM model from the Planck CMB observation are rather tight; for example, the angular acoustic scale is measured to 0.03\% precision, and other parameters are all measured to less than 1\% precisions except for the reionization optical depth $\tau$ (that is only measured to about 13\% precision). 

% The model gives a simple recipe for the cosmos: . It can easily yield a reasonable expansion history of the universe, and more importantly, it can reproduce statistically, through computer simulations, the properties of the cosmic microwave background (CMB) and the distribution of galaxies. 

%However, the standard model of cosmology currently cannot explicitly explain what the physical natures of dark energy and dark matter are, or how inflation happened. 
However, although the base-$\Lambda$CDM cosmology seems to be able to explain various cosmological observations quite well, some cracks still appear in the model. 
Recently, it was found that the Planck results for the base-$\Lambda$CDM cosmology are in tension with some low-redshift observations. For example, the amplitude of the fluctuation spectrum in the base-$\Lambda$CDM cosmology, constrained by the Planck observation, is found to be higher than that inferred from some analyses of cluster counts, weak gravitational lensing, and galaxy clustering; more importantly, the Planck base-$\Lambda$CDM results are in significant, 4--6$\sigma$, tension with local measurements of the Hubble constant (which prefer a higher value) \cite{Riess:2020sih}. 
These discrepancies in observations imply that the current standard model of cosmology is tantalizingly incomplete,
and needs to be extended. 
Theorists have designed a variety of wild theories beyond the standard model, introducing new physics and additional parameters into the extended models.

%and actually, theorists have designed all sorts of wild theories beyond the standard model. The possible, typical new physical elements may include, e.g., a dynamical dark energy, a direct interaction between dark energy and dark matter, a ``fifth force'' induced by some modified gravity theories, massive active/sterile neutrinos, extra relativistic species, a non-zero spatial curvature, primordial gravitational waves, and so forth. Considering these new physics needs to introduce new, additional parameters into the extended models. 

Nevertheless, the CMB observation is a two-dimensional measurement of the early universe, and thus cannot provide tight 
constraints on the parameters that concern the physical effects of the late universe. The CMB-alone constraints will always 
lead to significant degeneracies between cosmological parameters (including, in particular, the newly introduced parameters) in the extended cosmological models. Therefore, late-universe (low-redshift) astrophysical observations are needed to be combined with the Planck CMB observation to break the parameter degeneracies.
The most typical low-redshift cosmological probes include the type Ia supernovae (SN) and the baryon acoustic oscillations (BAO). The SN observation can provide cosmological ``standard candles'', and the BAO observation can provide a cosmological ``standard ruler''. Using the standard candles and the standard ruler, one can establish the distance--redshift relation for the cosmos, which embodies the expansion history of the universe, and thus can be used to constrain the properties of dark energy. In addition, other important cosmological probes also include the measurements for the Hubble constant, redshift-space distortions (RSD), weak lensing (WL), and clusters of galaxies. The RSD, WL, and clusters observations mainly aim to measure the structure growth history of the universe, and thus are important for the tests of the modified gravity theories (in which the predicted linear growth of structure is usually scale-dependent, fairly different from the prediction in General Relativity).
In the next one to two decades, the Stage IV dark energy experiments, such as DESI, LSST, Euclid, and WFIRST, 
will greatly improve the measurements for the parameters of dark energy and other cosmological parameters,
by performing optical or near-infrared sky surveys. 
%(with spectroscopic or imaging methods), by which the traditional cosmological probes (SN, BAO, RSD, and WL) will be greatly developed. 

Notwithstanding, with the great advancements in electronics, engineering, and computer science,
new cosmological probes are being developed, in addition to those traditional galaxy surveys.
In our opinion, the most important new cosmological probes mainly include gravitational-wave (GW) standard siren observation and neutral hydrogen (HI) 21 centimeter radio observation.
The future GW observations will play a crucial role in the studies of cosmology \cite{Zhang:2019ylr}.  Meanwhile, radio observations of the HI 21 cm signal from different cosmic epochs will also become a new powerful cosmological probe in the future. With the cosmological 21 cm signal as one of the main science drivers, the ambitious big-science facility, the Square Kilometre Array (SKA), is scheduled to be constructed in the near future.

%The GW multi-messenger observations can be used to study cosmology, and the sources of GWs in this type of observations are also called cosmological ``standard sirens'', by which a true distance--redshift relation can be established. The future GW detectors, such as the third-generation ground-based detectors, e.g., Einstein Telescope (ET) and Cosmic Explorer (CE), and the space-based detectors, e.g., LISA, Taiji, and TianQin, will definitely observe a large amount of GW events, and thus the GW standard sirens will certainly be developed into a new cosmological probe in the future. We refer the reader to Ref.~\cite{Zhang:2019ylr} for a brief review.

\begin{figure*}[!htp]
\includegraphics[scale=0.25]{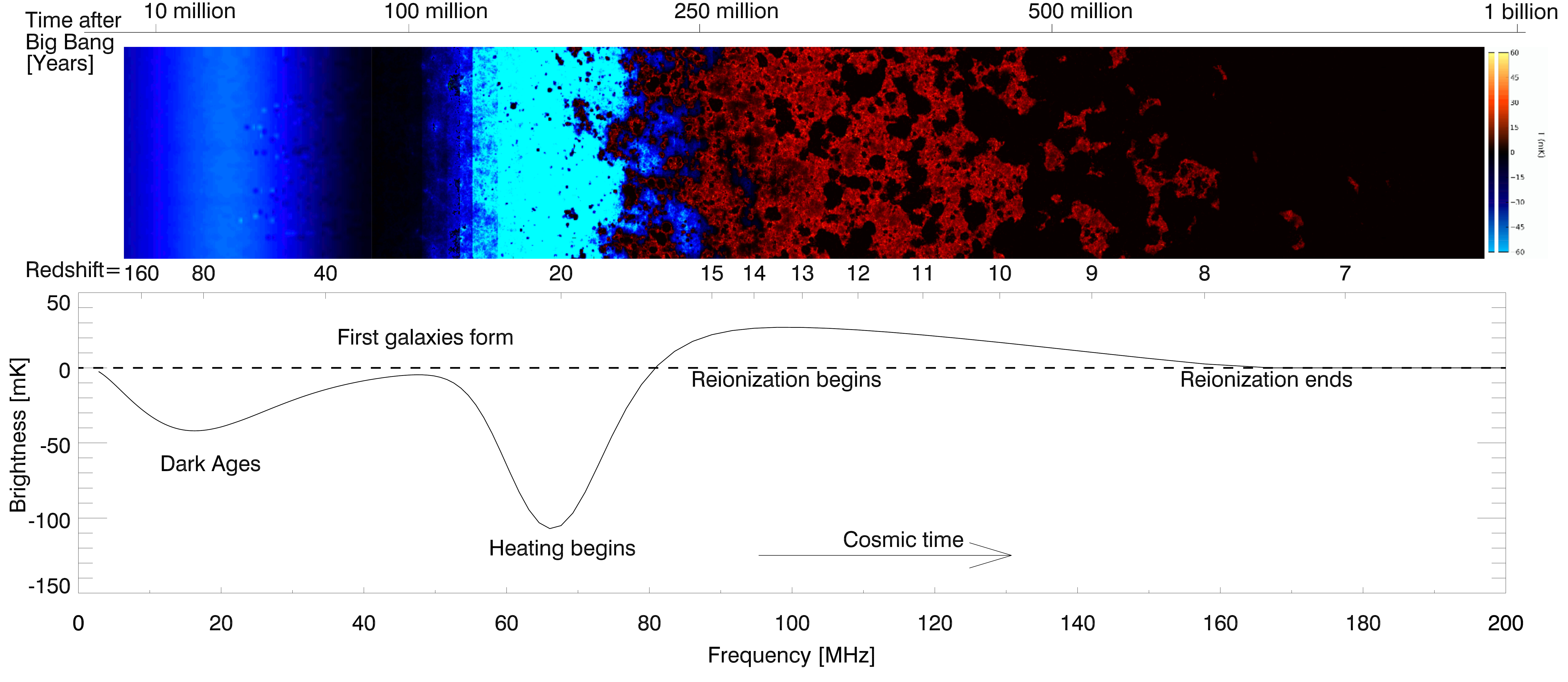}
\centering
\caption{\label{fig1} The cosmic 21 cm signal. {\it Upper panel:} Time evolution of fluctuations in the 21 cm brightness from the ``dark ages'' through to the end of the reionization epoch. This evolution is pieced together from redshift slices through a simulated cosmic volume. Coloration indicates the strength of the 21 cm brightness as it evolves through two absorption phases (blue), separated by a period (black) where the excitation temperature of the 21 cm hydrogen transition decouples from the temperature of the hydrogen gas, before it transits to emission (red) and finally disappears (black) owing to the ionization of the hydrogen gas. {\it Lower panel:} Expected evolution of the sky-averaged 21 cm brightness from the ``dark ages'' at redshift 200 to the end of reionization, around redshift 6 (the solid curve indicates the signal, and the dashed curve indicates zero-differential brightness). The frequency structure within this redshift range is driven by several physical processes, including the formation of the first galaxies and the heating and ionization of the hydrogen gas. There is considerable uncertainty in the exact form of this signal, arising from the unknown properties of the first galaxies. This figure is taken from Ref.~\cite{Pritchard:2011xb}. } 
\end{figure*}

%The SKA is an international project of the next-generation radio observatory that will be built by more than 10 countries. It will ultimately have an effective collecting area of about $10^6$ m$^2$, i.e., the collecting area necessary to detect the HI emission at 21 cm from an $L_\ast$ galaxy at $z\sim 1$ in a few hours. 

The SKA will comprise of two arrays, i.e., a mid-frequency dish array (SKA-MID) to be built in South Africa, and a low-frequency array of log-periodic antenna elements (SKA-LOW) to be built in Australia, totally covering the frequency range from 50 MHz to 20 GHz. It will be built in two stages. The phase 1 of SKA (denoted as SKA1) will begin construction in 2021 (lasts from 2021 to 2028), and the phase 2 of SKA (denoted as SKA2) will start construction after 2028.

The primary science objective of the SKA is to explore the first billion years of the universe, with a special focus on the cosmic dawn (CD) and the epoch of reionization (EoR). Hydrogen is ubiquitous in the universe, and it constitutes about 75\% of the gas mass in the intergalactic medium (IGM), thus it actually provides a convenient tracer of the properties of the gas before the completion of reionization. The HI 21 cm line is produced by the hyperfine splitting caused by the interaction between the magnetic moments of electron and proton in the hydrogen atom. 
The 21 cm signals from HI during CD and EoR are redshifted to the low-frequency radio band today.
The key features of the theoretical HI 21 cm signal from the first billion years of the universe are shown in Fig.~\ref{fig1} (see also Ref.~\cite{Pritchard:2011xb}). The upper panel shows the time evolution of fluctuations in the 21 cm brightness, and the lower panel shows the expected evolution of the sky-averaged 21 cm brightness.

The SKA seeks to explore the CD and EoR by performing the imaging and power spectrum measurements of the 21 cm brightness field. The SKA1-LOW array can be used for 21 cm signal detection in the frequency range of 50--200 MHz to probe the physics of CD and EoR at the redshifts 6--27. It is expected that the first image of the cosmic reionization in the frequency range of 100--200 MHz could be obtained, which can be used to uncover the mysterious veil of the cosmic reionization, and the HI power spectra of CD and EoR on the scales 0.02--0.1 Mpc$^{-1}$ in the frequency range of 50--200 MHz could also be statistically obtained. EoR imaging observations can reveal the formation of the first stars and black holes as well as the complex process of gas reionization.

%, reproducing the universe's history from darkness to brightness. Compared with the SKA pathfinder projects (such as MWA, LOFAR, PAPER, and 21CMA) currently in operation, the SKA1-LOW array is the only experimental instrument that can realize EoR imaging observation, which is of great scientific significance. 

The EoR imaging observations are just one of many scientific goals of the SKA. The five major scientific goals of the SKA project include: (i) CD and EoR explorations; (ii) galaxy evolution, cosmology, and dark energy; (iii) the cradle of life; (iv) fundamental physics with pulsars; and (v) origin and evolution of cosmic magnetic field. 

%These five scientific goals are further refined into 16 research directions. The SKA is a big scientific instrument that will ultimately answer some of the most fundamental questions of the universe. 

%and with the answers to these fundamental questions will open up a new era in our understanding of the universe.

\begin{figure*}[!htp]
\includegraphics[scale=0.4]{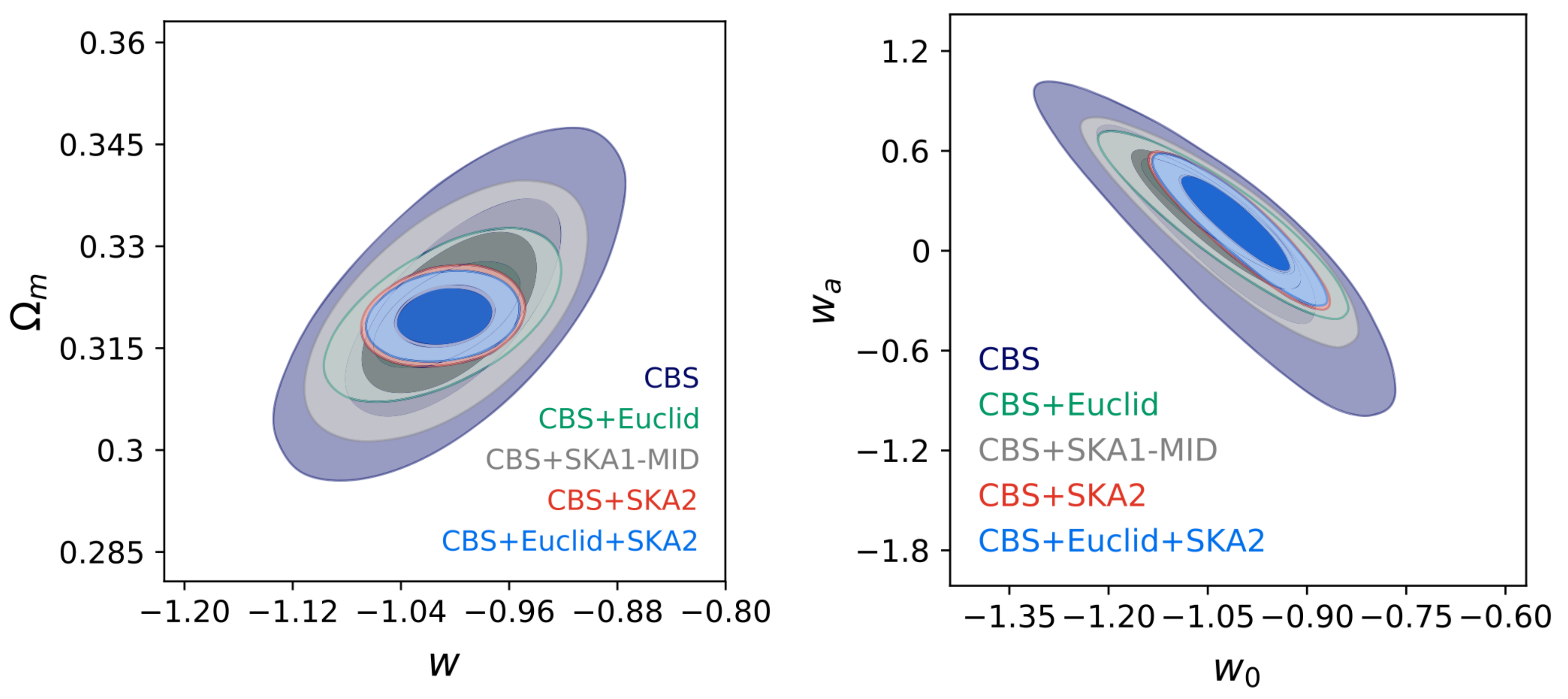}
\centering
\caption{\label{fig2} Observational constraints (68.3\% and 95.4\% confidence level) on the $w$CDM ({\it left}) and $w_0w_a$CDM ({\it right}) models by using the CBS, CBS + Euclid, CBS + SKA1, CBS + SKA2 and CBS + Euclid + SKA2 data combinations. Here, CBS stands for CMB + BAO + SN. This figure is taken and adapted from Ref.~\cite{Zhang:2019dyq}.}
\end{figure*}

Actually, by using the 21 cm signal of HI as a tracer of the matter distribution, the 21 cm cosmology is also one of the 
breakthroughs the SKA is attempting to make. 
In addition to the SKA-LOW for the early structure formation studies, the design of SKA-MID array has a special focus on 
the evolution of the late universe and the nature of dark energy.
The SKA1-MID will be a mixed dish array consisting of an existing SKA precursor, MeerKAT, of 64 13.5-m diameter dishes, and a newly-built array of 133 15-m dishes, covering the frequency range from 350 MHz to 14 GHz. These dishes are planned to be equipped with receivers sensitive to 5 different frequency ranges (or {\it bands}), but in the present SKA baseline configuration there are only sufficient funds to deploy Band 1 (0.35--1.05 GHz) and Band 2 (0.95--1.75 GHz), which are most relevant to cosmology, and Band 5 \cite{RedBook2018}.

The SKA-MID array can be used to implement the HI sky survey and study the large-scale structure of neutral hydrogen. An effective way to map the sky called the ``intensity mapping'' (IM) adopts a low angular resolution approach (in which each pixel contains many galaxies) to perform the sky survey, and thus can get the large-scale distribution of neutral hydrogen with extremely high efficiency. The 21 cm IM survey by the SKA could provide accurate measurements for the HI power spectrum, and the associated signals of BAO and RSD can reach percent-level precision in the expansion rate measurements of the universe, providing effective constraints on the equation of state (EoS) of dark energy
(as well as relevant parameters of the modified gravity models) up to $z\sim3$.
Besides dark energy, the SKA-MID survey could also make important tests %on the General Relativity, and 
on the isotropy, homogeneity, and non-Gaussianity of the universe.
%China has a good foundation in the field of neutral hydrogen survey and the relevant cosmological research, and has independently built the SKA pathfinder, the Tianlai array, and the world's largest single-dish radio telescope, the FAST. With these two experimental instruments (an interferometer array and a single-aperture telescope, respectively), we have accumulated a wealth of experience and trained a number of researchers in the related fields, as well as conducted a large amount of preliminary researches on SKA's neutral hydrogen survey and 21 cm cosmology (see, e.g., Refs.~\cite{Xu:2017rfo,Yohana:2019ahg,Zhang:2019dyq,Zhang:2019ipd,Liu:2019asq,Jin:2020hmc,Xu:2016kwz,Xu:2014bya,Huang:2018ral,antao}). 
Some relevant studies \cite{Xu:2017rfo,Yohana:2019ahg,Zhang:2019dyq,Zhang:2019ipd,Liu:2019asq,Jin:2020hmc,Xu:2016kwz} have shown that the SKA-MID HI 21 cm IM survey could play an important role in improving the cosmological parameter estimation in the future.

%It has been shown \cite{Zhang:2019dyq,Xu:2017rfo,Zhang:2019ipd,Xu:2016kwz} that the SKA-MID HI 21 cm IM survey could play an important role in improving the cosmological parameter estimation in the future, in particular for the constraints on the EoS of dark energy \cite{Zhang:2019dyq}, the possible interaction between dark energy and dark matter \cite{Xu:2017rfo}, the total mass of neutrinos \cite{Zhang:2019ipd}, and the primordial non-Gaussianity and inflation features \cite{Xu:2016kwz}. It can be expected that in the future the SKA observation, combined with the future highly accurate optical surveys from the Stage IV dark energy experiments, such as DESI, LSST, and Euclid, as well as the gravitational-wave standard siren observations from ground-based and space-based detectors, would greatly promote the development of cosmology.

%(see, e.g., Refs.~\cite{Xu:2017rfo,Yohana:2019ahg,Zhang:2019dyq,Zhang:2019ipd,Liu:2019asq,Jin:2020hmc,Xu:2016kwz,Xu:2014bya,Huang:2018ral,antao}). 

For example, it is shown in a recent work \cite{Zhang:2019dyq} that, compared to the currently best constraints by the CMB+BAO+SN data (hereafter abbreviated as CBS for convenience), the SKA1 data can improve the constraints on $\Omega_{\rm m}$ and $H_0$ by about 20\%, and SKA2 data can improve the constraints by about 60--70\%. Although the cosmological parameter constraint capability of SKA1 is weaker than that of the Stage IV dark energy optical surveys, e.g., the Euclid, the SKA2 is much better than them. For the EoS of dark energy, in the $w$CDM model, $w$ will be measured to be $\sigma(w)\approx 0.04$ by CBS+SKA1 and $\sigma(w)\approx 0.02$ by CBS+SKA2; in the $w_0w_a$CDM model, $w_0$ and $w_a$ will be measured to be $\sigma(w_0)\approx 0.08$ and $\sigma(w_a)\approx 0.25$ by CBS+SKA1, and $\sigma(w_0)\approx 0.05$ and $\sigma(w_a)\approx 0.18$ by CBS+SKA2. The constraints on the $w$CDM and $w_0w_a$CDM models (in the $w$--$\Omega_{\rm m}$ and $w_0$--$w_a$ planes, respectively) by using the CBS, CBS+Euclid, CBS+SKA1, CBS+SKA2, and CBS+Euclid+SKA2 data combinations are shown in Fig.~\ref{fig2}. Note that in this study the forecasted BAO measurements by SKA1-MID (IM) and SKA2 (galaxy survey) \cite{Bull:2015nra} are used. For the Fisher forecasting formalism developed for the HI 21 cm survey, we refer the reader to Ref.~\cite{Bull:2014rha}. See also, e.g., Refs.~\cite{Xu:2017rfo,Yohana:2019ahg}, for other relevant studies on constraining dark energy models with the forecasted SKA-MID data. Here we also note that the synergy of HI 21 cm observation and GW standard siren observation has been discussed in Ref.~\cite{Jin:2020hmc} (exampled with the synergy of SKA and CE), and the constraints on dark energy with the redshift-drift observation of SKA have been detailed in Ref.~\cite{Liu:2019asq}.

The SKA-MID data can also be used to constrain the neutrino mass. In Ref.~\cite{Zhang:2019ipd}, the issue of using the SKA data to improve the constraints on the total neutrino mass is discussed. This work does not assume a detection of the neutrino mass in the cosmological fit, but only considers upper limits on the neutrino mass. It should be noted that, although the total neutrino mass has still not been determined, the upper limits on it have been rather stringent using the current CBS data. In Ref.~\cite{Zhang:2019ipd}, the three cases of the neutrino mass ordering are separately discussed, and it is shown that using the future 21 cm observations from the SKA, the upper limits on the total neutrino mass can be further decreased by about 4\% (NH), 3\% (IH), and 10\% (DH) for SKA1, and by about 7\% (NH), 7\% (IH), and 16\% (DH) for SKA2. Here, NH, IH, and DH stand for the normal hierarchy, the inverted hierarchy, and the degenerate hierarchy of the neutrino mass ordering, respectively. The forecasted constraints on the neutrino mass by SKA1-MID plus SKA1-LOW by assuming a detection of the neutrino mass in cosmology can be found in Ref.~\cite{RedBook2018}.

\begin{figure}[!htp]
\includegraphics[scale=0.35]{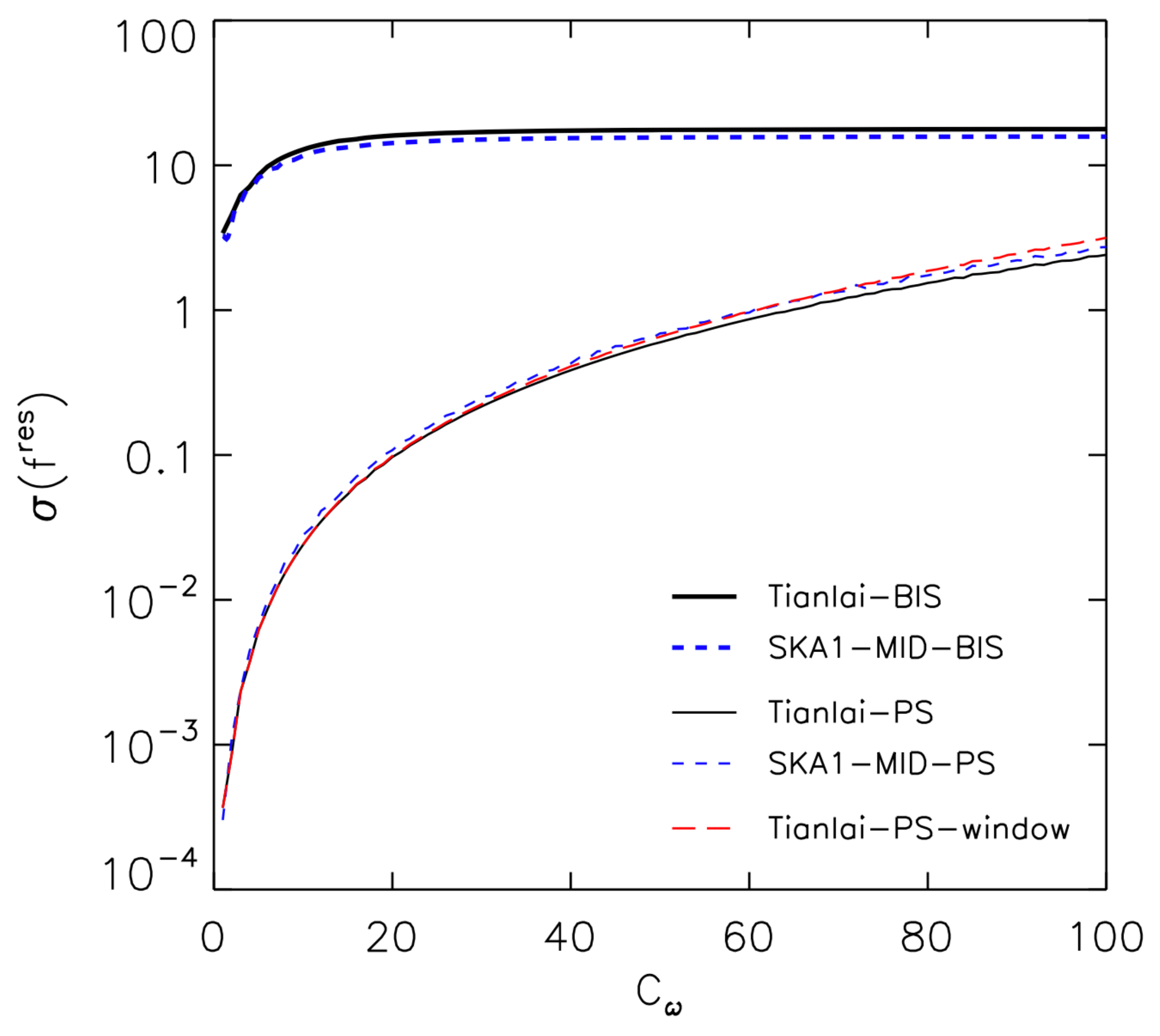}
\centering
\caption{\label{fig3} The marginalized 1$\sigma$ error on $f^{\rm res}$ as a function of $C_\omega$  in the resonant model for HI power spectrum measurements (thin lines) and for HI bispectrum measurements (thick lines), with the fiducial value of $f^{\rm res}$ set to zero. In each set of lines, the solid and dashed lines are for Tianlai and SKA1-MID, respectively. The thin long-dashed line shows the HI power spectrum measurement with Tianlai when the window function effect is taken into account. This figure is taken from Ref.~\cite{Xu:2016kwz}.}
\end{figure}

In addition to the dark energy parameters and the neutrino mass, the SKA-MID HI 21 cm IM survey can also provide hints on the primordial non-Gaussianity (PNG). It is predicted that making use of the effect of scale-dependent bias, we can achieve $\sigma(f_{\rm NL}) = 2.8$ with the SKA1-MID, thanks to the extremely large survey area achievable with the IM technique \cite{RedBook2018}. Any constraint on the PNG could shed light on the physics of inflation. Especially, as highlighted in Ref.~\cite{Xu:2016kwz}, using the 21 cm IM survey for exploring a special class of inflation models, which imprint features in the primordial power spectrum and bispectrum, is very promising, because of the three-dimensional information retained in the 21 cm IM survey, as compared to the two-dimensional CMB observations, and the ultra-large survey volumes accessible by the SKA-MID. The potential of such surveys for detecting the inflationary ``resonant'', ``kink'', ``step'', and ``warp'' features in the observable power spectrum and bispectrum has been discussed in Refs.~\cite{RedBook2018,Xu:2016kwz}. Figure~\ref{fig3} shows the forecasted constraints on the amplitude of the resonant PNG, $f^{\rm res}$, as a function of the resonance frequency $C_{\omega}$, using either the scale-dependent bias of the power spectrum or the bispectrum, with the Wide Band 1 Survey of SKA1-MID (adding Band 2 IM observations for $z<0.4$). Here, the results from the Tianlai observation are also shown for comparison (see also Ref.~\cite{Xu:2014bya} for other forecasted results of the Tianlai). It is found that even in the presence of the foreground contamination, the upcoming 21 cm IM observations of the large-scale structure in the post-reionization universe alone could put extremely tight constraints on the feature models, potentially achieving orders-of-magnitude improvements over the two-dimensional CMB measurements.

%Here we do not wish to give a complete review on the forecasting parameter constraint research of the 21 cm cosmology based on SKA, but we only wish to show by taking several typical examples that the SKA 21 cm observations could play a crucial role in improving cosmological parameter estimation in the future. 

It has been shown \cite{Zhang:2019dyq,Xu:2017rfo,Zhang:2019ipd,Xu:2016kwz} that the SKA-MID HI 21 cm IM survey could play an important role in improving the cosmological parameter estimation in the future, in particular for the constraints on the EoS of dark energy \cite{Zhang:2019dyq}, the possible interaction between dark energy and dark matter \cite{Xu:2017rfo}, the total mass of neutrinos \cite{Zhang:2019ipd}, and the primordial non-Gaussianity and inflation features \cite{Xu:2016kwz}. It can be expected that in the future the SKA observation, combined with the future highly accurate optical surveys from the Stage IV dark energy experiments, such as DESI, LSST, and Euclid, as well as the gravitational-wave standard siren observations from ground-based and space-based detectors, would greatly promote the development of cosmology.

\end{document}